\documentstyle[twoside]{article}

\begin{document}
\begin{center}
{\Large \bf From charge transfer type insulator to
superconductor}\\

\end{center}
\begin{center}
{\normalsize  Wei Guo, Rushan Han }
\\
{\it School of Physics, Peking University, Beijing 100871 }
\end{center}
${}$\\
\begin{center}\begin{minipage}{4.5in}
{\footnotesize We propose a microscopic model Hamiltonian to
account for impurity doping induced insulator-superconductor
transition and the coexistence of antiferromagnetism and
superconductivity in the high-$T_c$ cuprates. The crossover from
non Fermi liquid to Fermi liquid regime characterized by
delocalization of $d$ electrons on Cu sites is discussed.}
\end{minipage}\end{center}

\section{Introduction}
The insulator-superconductor transition driven by impurity doping
occurs uniquely in a class of copper oxides with a special
magnetic structure. The magnetism of the high-$T_c$ cuprate arises
from unfilled $d$ shell ($3d^9$) of $\rm Cu^{2+}$ ions in the $\rm
CuO_2$ plane, the overlap of the atomic $d$ wave function is weak
because of large spacing between Cu ions($3.8\AA$). The strong
correlation between local Cu moments($S=1/2$) comes from the
indirect interaction mediated by $p$ electrons at neighbor oxygen
sites, i.e. the superexchange$[1][2]$. Even though the superfluid
carriers in the high-$T_c$ cuprates have the $d$ wave character,
however experimental observations confirm that the doped holes
actually appear in the $p$ band$[3]$, which is consistent with the
superexchange picture since $d$ electrons are localized in the
doping range $0 < x < 0.2$. In this article we show that $d$ wave
superconductivity is not originated from hopping $d$ electrons but
$p$ electrons. Undoped high-$T_c$ cuprates are in fact charge
transfer type insulators$[4]$, where $p$ band is above lower $d$
band. $p$ electrons being the mediation for magnetic ordering in
the insulating phase became conductive after impurity doping,
which causes the partial occupancy of the Wannier $p$ state at a
portion of oxygen sites and local $d$-level shift at neighbor Cu
sites. If local $d$-level merge into $p$ band, the hybridization
between $p$ and $d$ electrons generates the Kondo exchange
interaction between itinerant $p$ electron spins and the local Cu
spin. We show that at such a Cu site, the local Cu spin can be a
mediation for electron pairing via Kondo exchange. The effective
electron-electron interaction originated from the spin coupling is
{\it attractive } and may yield a bound spin triplet with opposite
spins $\chi_+ = |\uparrow \downarrow \:\rangle + |\downarrow
\uparrow \:\rangle $($S=1, S_z=0$). The $d$ wave symmetry of
paired electrons comes from $l=2$ partial wave channels of Bloch
states forced by crystal and spin pair symmetry. Since electron
pairs couple to the local spins in the $\rm CuO_2$ plane the
effective {\it pair-pair }potential is strong and leads to high
temperature superfluid transition.
\section{Hamiltonian}
The high-$T_c$ cuprates are essentially interacting spin systems.
In the low doping limit, its main electronic structure consists of
localized $d$ electrons and conductive $p$ electrons. To begin our
analysis, we first divide Cu spins in the $\rm CuO_2 $ plane into
clusters. A spin cluster consists of a Cu spin at the center of
the cluster and four nearest Cu spins. Interactions between $d$
electron at the central Cu site and $p$ band electrons can be
written in terms
\begin{equation}
\hat H^\prime =\sum_{\vec k \sigma } \epsilon_{\vec k } \hat
c_{\vec k \sigma }^\dagger \hat c_{\vec k \sigma } + \sum_{\sigma
} \epsilon_{d } \hat d_{\sigma}^\dagger \hat d_{\sigma } +
\sum_{\vec k \sigma }( V_{\vec k }\hat d_{\sigma }^\dagger\hat
c_{\vec k \sigma } + V_{\vec k }^{*} \hat c_{\vec k \sigma
}^\dagger \hat d_{\sigma })+ U_d \hat n_{d \uparrow}\hat n_{d
\downarrow } + \frac 1 N \sum_i U_p \hat n_{p\uparrow }^i \hat
n_{p \downarrow }^i
\end{equation}
where $\hat c_{\vec k \sigma }^\dagger $ is the creation operator
of the Bloch state $\phi_{\vec k \sigma }(\vec r )$ from $p$ band.
$\epsilon_d $ is the local $d$ level, $V_{\vec k }$ is the mixing
between $d$ and $p$ electrons. The on-site Coulomb interaction
$U_d$ and $U_p$ at copper and oxygen sites are considered, $\hat
n_{d\sigma }$ and $\hat n_{p\sigma }^i $ are occupancy operators
for Wannier states $\phi_{d\sigma }$ and $\phi_{p\sigma }^i$, $N$
is the total number of oxygen sites. By using mean field
approximation, the term $U_p\hat n_{p\uparrow }^i \hat
n_{p\downarrow }^i $ in Eq.(1) can be written as $U_p (\hat n_{p
\uparrow }^i \langle \hat n_{p \downarrow }^i \rangle + \hat n_{p
\downarrow }^i \langle \hat n_{p \uparrow }^i \rangle - \langle
\hat n_{p \uparrow }^i\rangle \langle \hat n_{p\downarrow }^i
\rangle )$, we can expand $\hat n_{p\sigma }^i$ in the Bloch
representation
\begin{equation}
\hat n_{p\sigma }^i = \sum_{\vec k } \hat n_{\vec k \sigma }e^{i
\vec k \cdot \vec R_i }
\end{equation}
the mean field term can be added to $\epsilon_{\vec k }$ in
Eq.(1), the normalized $p$ band energy is $\overline
\epsilon_{\vec k \sigma }= \epsilon_{\vec k } + U_0$, $\hat
H^\prime $ has the same form as the Anderson impurity model. We
can write $U_0$ in the average form for the nearest-neighbors
\begin{equation}
U_0 = \frac 1 N_s \sum_i U_p\langle \hat n_{p-\sigma }^i \rangle
\end{equation}
sum $\sum_i$ is over the nearest oxygen sites($N_s$). For undoped
compounds $\langle n_{p-\sigma }^i \rangle =1 $, the $p$ band is
above local $d$ level for large $U_0$ and filled. In this case,
the $p$ band and the local $d$ level is separated by a energy gap,
the mixing interaction $V_{\vec k } = 0 $. For doped compounds,
the doped holes may appear in the nearest oxygen sites of a Cu
site, the partial occupancy of $p$ electron reduces the strength
of $U_0$ and causes local energy levels shift. If local $d$ level
at a Cu site merge into the $p$ band, the hybridization of $p$ and
$d$ electrons $V_{\vec k }$ generates Kondo exchange interaction
since the Anderson model is equivalent to the Kondo exchange model
by a canonical transformation$[5]$. Therefore, we can derive a
$p$-$d$ exchange term from the mixing interaction in Eq.(1), which
is
\begin{equation}
\hat H_{ p-d } =- J{\bf S_0 \cdot \sigma }(0)
\end{equation}
where $\sigma (0)$ is the spin density of electrons at$ \vec R_0$.
The criterion for localization of a Cu spin in the metallic
environment is given by
\begin{equation}
\langle \hat n_{d\sigma } \rangle = \frac 1 2 - \frac 1 \pi
\tan^{-1} \left ( \frac {\overline \epsilon_{d \sigma }-
\epsilon_F } \Delta \right )
\end{equation}
$\overline \epsilon_{d \sigma }$ is the normalized local $d$
level, $\Delta =\pi \sum_{\vec k }|V_{\vec k }|^2 \delta (\epsilon
-\epsilon_F )$$[6]$. The localization of $d$ electron requires
$\epsilon_F - \overline \epsilon_{d \sigma }>> \Delta $, $d$ level
is far below the Fermi energy in the low doping limit.

The next nearest-neighbors of spin $\bf S_0$ are four Cu spins
$\bf S_j$, the indirect interaction between Cu spins via $p$-$d$
coupling can be expressed in terms of the superexchange
\begin{equation}
\hat H^{\prime \prime } = \frac K 2 \sum_j {\bf S_0\cdot S_j }
\end{equation}
sum $\sum_j$ is over the nearest Cu sites of $S_0$, the factor 1/2
is taken from the fact that $p$-$d$ transition occurs at one end
of CuO bond, direct exchange at the other end. In a square lattice
we have $two$ superexchange bond instead of four. The
superexchange arises from $p$-$d$ coupling, the coupling constant
$K =\rho^2 J_d $, where $J_d$ is the coupling constant of direct
exchange, $\rho$ the transition integral
\begin{equation}
\rho = \int d^3 x \phi_p ({\bf x })V\phi_d ({\bf x})
\end{equation}
The superexchange in undoped compounds is strong, $K_0 = 0.1 \rm
eV$. If the doped holes appear in the nearest-neighbor of $S_0$,
the superexchange as well as $U_0$ are suppressed by holes easing
the restriction for local Cu spin flip. On the other hand, the CuO
bond of spin $\bf S_j$ on the boundary of the cluster is less
affected by holes, $U_0$ keeps large at site $j$, there is no
mixing between $d$ and $p$ electrons at such sites, there is one
Kondo scatter per spin cluster on the average, the spin cluster
model is different from Kondo lattice model. $\hat H^{\prime
\prime }$ can be further simplified by considering
\begin{equation}
\hat H^{\prime \prime }  = \frac K 2 \sum_j [ S_0^z S_j^z + \frac
1 2 ( S_0^+S_j^- + S_0^-S_j^+ )]
\end{equation}
the last two terms in Eq.(8) flips $S_j$, which is a higher energy
process since $S_j$ is stiff in the robust CuO bonds. These two
terms can be ignored as far as low energy excitation is concerned.
In a proper doping range, we have the total Hamiltonian $ \hat H =
\hat H^\prime + \hat H^{\prime\prime }$ by considering
interactions from the nearest and the next nearest-neighbors of a
local Cu spin in a spin cluster
\begin{equation}
\hat H  = \sum_{\vec k \sigma } {\overline \epsilon}_{\vec k
\sigma } \hat c_{\vec k \sigma }^\dagger \hat c_{\vec k \sigma } +
\frac K 2 \sum_j  S_0 S_j - J{\bf S_0 \cdot \sigma }(0)
\end{equation}
where $S_0$,$S_j =\pm 1 $. The anisotropic Ising-type local spin
correlation conforms to the experimental result for spin
susceptibility measurement$[7]$. Magnetic ordering of local spins
is a key factor to the pairing mechanism. The existence of $p$-$d$
exchange interaction is justified by the observed anomalous Hall
effect in the normal states of high-$T_c$ cuprates, which is a
common feature of Kondo-type systems originated from $l$ partial
wave of conduction electrons and local moment interaction
\begin{equation}
\hat H_l \sim {\bf l \cdot S_0 }
\end{equation}
The partial wave coupling and its contribution to anomalous Hall
effect was calculated by using the Anderson model$[8]$. This type
of interaction is irrelevant to electron pairing, we ignore it in
the derivation of Eq.(9).
\section{Spin Paring}
From Eq.(9) we can show the instability of the normal states of
$p$ band electrons caused by spin coupling. We consider two
electrons interact with the local spin $S_0$ via the Kondo
exchange, which is
\begin{equation}
\begin{array}{lcr}
{\hat H_{int} = -J {\bf S_0 \cdot ( \sigma_1  + \sigma_2  )}} &&
(J < 0)
\end{array}
\end{equation}
we calculate the energy variation to second order
\begin{equation}
\delta E = \langle 0|\hat H_{int}|0 \rangle + \frac { \langle
0|\hat H_{int}|1 \rangle \langle 1|\hat H_{int}|0 \rangle }{ E_0 -
E_1 }
\end{equation}
where $|0\rangle $ is the ground state,$|1\rangle $ the excited
state. The local spin flip can be a second order spin interaction
effect rather than quantum fluctuation. From Eq.(12) we obtain an
effective interaction between two electrons, which is
\begin{equation}
\hat H_{ ef} = - \frac 1 4 \lambda J (\sigma_{1}^- + \sigma_{2}^-
)(\sigma_{1}^+ + \sigma_{2}^+ ),
\end{equation}
where $\lambda =J/2K $, $J$ is in order of $10^{-2}\rm eV$$[8]$.
$\lambda $ can be used as a small parameter in calculation of the
bind energy of electron pairs.  It can be shown that $\hat H_{ef}$
has a bound state solution, which is a spin triplet with opposite
spins
\begin{equation}
\chi_+ = |\uparrow \downarrow \:\rangle + |\downarrow \uparrow
\:\rangle
\end{equation}
with binding energy
\begin{equation}
E_b = -\frac 1 2 \lambda J
\end{equation}
the energy of pair states $|\uparrow \uparrow \: \rangle $ and
$|\downarrow \downarrow \: \rangle $ depends on $\langle
\,0|S_i|0\,\rangle $, which violates the translational invariance
in the presence of antiferromagnetic background, therefore
parallel spin pairs are unphysical states. It is easy to show that
the spin singlet is an excited state. The formation of the
unconventional electron pair requires local moment $S=1/2$ that
forbids non-copper oxides being unconventional superconductors.
The intrinsic inhomogenity of electronic states in atomic scale is
an indication of local pairing, which is observed by using STM
technique$[9]$.
\section{D-Wave Paring }
Local triplet pairing excludes $s$ wave pairing, the wave function
of electron pair must have $d$ wave symmetry in restrictions of
tetragonal symmetry of crystal lattice. So that Bloch electrons
may pair through $l=2$ partial wave channel by taking $m=\pm 1$
\begin{equation}
{\psi} ( x_1 , x_2 ) \sim \sin {\theta}_1 \cos {\theta}_1 \sin
{\theta }_2 \cos {\theta }_2  \left ( \exp i  ({\varphi}_1 -
{\varphi}_2 ) - \exp i ( {\varphi}_2 - {\varphi}_1 ) \right )
\end{equation}
the partial wave states may give a coherent length in order of
$10\AA $, which has been measured in experiments and is difficult
to explain by using terms of atomic Wannier wave function. Since
electron motion is limited to the layered $\rm CuO_2 $ planes,
what we observed is actually its two dimensional projection( with
z axis in the plane )
\begin{equation}
\begin{array}{lcr}
{\psi_\theta \sim - \cos 2 \theta } && { (-\pi \leq \theta \leq
\pi ) }
\end{array}
\end{equation}
where $\theta = \theta_1 - \theta_2 $ is the relative coordinate
of two electrons, which is invariant in space reflection. However,
the three dimensional wave function itself is antisymmetric, so
that Eq.(17) is consistent with the spin triplet state. It is
difficult to distinguish a singlet and a triplet state with
opposite spins in experiments,  a convincing experimental evidence
for identifying the spin state of the electron pair in high-$T_c$
cuparates is still lack. The magnetic pair with $S=1$ breaks
time-reversal symmetry when it interacts with external probes.
Experimental observation of time-reversal symmetry breaking effect
in the high-$T_c$ cuprates has been reported by a research group
by using circularly polarized photon technique$[10]$, which we
consider is an indication for the existence of magnetic pair
state.
\section{ Mean Field Theory }
By using mean field theory to calculate the superfluid transition
temperature we obtained a BCS-like expression
\begin{equation}
k_B T_c = \omega_0 \exp( \frac 1 {N(0)V })
\end{equation}
where $\omega_0 =|E_b|$, the character energy scale for magnetic
coupling, $N(0)$ is the density of states at the Fermi
surface$[11]$. Since electron pairs couples to the whole $\rm
CuO_2$ plane, the effective pair potential $V$ is strong. Strongly
coupled pair-pair interaction may give rise to the pseudogap in
excitation spectrum and linear temperature dependence of
resistivity in the normal states$[12]$. To explain the electronic
phase diagram, we calculate the binding energy $E_b$ to fourth
order in $J$ and take the strong coupling limit$N(0)V >> 1$, we
obtain an expression for $T_c$
\begin{equation}
k_B T_c = \frac 1 2 \lambda J ( 1 - \frac 1 2 \lambda^2 )
\end{equation}
where $T_c$ increases with increasing hole concentration $x$ via
$\lambda $ until a maximum $T_c$ is reached
\begin{equation}
\lambda_m^2 = \frac 2 3
\end{equation}
Kondo coupling constant $J$ sets a natural limit on the maximum
$T_c$. By estimating $J$ up to $50\rm meV $, the upper limit of
$T_c$ is about $150K$. The universal expression for $T_c$ near
optimum doping can be derived from Eq.(19), which is
\begin{equation}
\frac {T_c} {T_m} = 1 - \kappa ( x - x_m )^2
\end{equation}
with
\begin{equation}
\kappa = (\frac {2K_0}{J})^2
\end{equation}
A analysis of experimental data for $\rm La_{2-{\it x } } Sr_{\it
x }CuO_4$ gives $\kappa = 92.4$, where we have taken $J= 26 \rm
meV $, $K_0 = 0.125 \rm eV$$[12]$. Theoretical value of $\kappa $
is close to the empirical value $82.6$$[13]$. $\kappa $ and $x_m$
are not independent parameters in spin pairing theory, we can show
the relation between them. $p$ electron wave function $\psi_p$ can
be written in terms
\begin{equation}
\psi_p = \mu |p\rangle + \nu |0\rangle
\end{equation}
where $ |p \rangle $ is the occupied state, $|0 \rangle $ is the
empty state, $\mu^2 $ and $\nu^2$ are probabilities for $p$
electron and hole occupancy satisfying $\mu^2 + \nu^2 = 1 $. The
strength of the superexchange is reduced by holes in a doped
high-$T_c$ cuprate $K= K_0 (1- \nu^2)$. At optimum doping $\nu_m$
is determined by
\begin{equation}
\lambda_m = \frac J {2K_0(1 - \nu_m^2 ) }
\end{equation}
By taking empirical value $\kappa = 82.6 $, we obtain $\nu_m^2
=0.85 $ from Eq.(24). If hole appears at two oxygen sites in a
spin cluster, the optimum impurity concentration is therefore $
x_m = 2\times 0.85/10 = 0.17 $. The positive charge of holes
balances the Coulomb repulsion of electrons pair which has a short
coherent length. It is easy to show from Eq.(15) that a portion of
$p$ electron becomes superfluid carriers, in the low doping limit
the superfluid density $n_s$ increases linearly with the doped
hole concentration $x$, thus we have
\begin{equation}
T_c \propto n_s \propto x
\end{equation}
The superexchange vanishes at a doping point where there are two
holes with occupancy $\nu^2 =1 $ residing in a spin cluster, the
average hole occupancy in the oxygen sites is 1/2. The condition
for charge balance at the central spin site requires $\langle \hat
n_{d,\sigma }\rangle =1/2 $. From Eq.(5), the local level shift
caused by impurity doping satisfies $\overline \epsilon_d -
\epsilon_F = 0 $, $d$ electron is delocalized at the doping
concentration $x = 0.2 $. The superconducting phase also vanishes
at this doping point. The delocalization of electrons also causes
a crossover from non Fermi liquid to Fermi liquid, since non Fermi
liquid behavior actually arises from electron-local-moment
interaction.
\section{Summary}
 In summary, we use a microscopic model Hamiltonian to describe
insulator- superconductor transition, the coexistence of
antiferromagnetism and superconductivity in high-$T_c$ cuprates.
By increasing impurity concentration we have the following
effects: i) metal-insulator transition, ii) Kondo scattering at a
portion of random Cu sites, iii) Ising-like short range magnetic
order, iv) spin pairing and the superconducting transition, v)
delocalization of $d$ electron and crossover from non Fermi liquid
to Fermi liquid regime. We have shown that $d$ wave
superconductivity is not originated from hopping $d$ electrons but
$d$ channels of Bloch electrons from $p$ band forced by spin
pairing. High temperature superfluid transition is not due to the
strong charge coupling between correlated electrons, but the weak
electronic correlation established in a strongly correlated
magnetic background. An external probe( circularly polarized
photons for example ) acting on a magnetic electron pair may break
time-reversal symmetry, which is the core prediction of spin
pairing theory, yet to be confirmed by experimentalists. As we
have shown, the concept of spin pairing is capable of achieving a
thorough comprehension of the origin of high-$T_c$
superconductivity.

\section*{References}
1. Kramer H A, 1934 Physica {\bf 1} 182\\
2. Anderson P W, 1950 Phys. Rev. {\bf 79 } 350\\
3. Romberg H, Alexander M, Nucker N, Adelmann, Fink J, 1990 Phys.
Rev. B {\bf 42 } 8768\\
4. Uchida S, Ido T, Takagi H, Arima T, Tokura Y, Tajima S,
Phys. Rev. B {\bf 43 } 7942\\
5. Schrieffer J R, Wolff P A, 1966 Phys. Rev. {\bf 149 } 491\\
6. Anderson P W, 1961 Phys. Rev. {\bf 124 } 41\\
7. Lavrov A N, Ando Y, Komiya S and Tsukada I, 2001 Phys. Rev.
Lett. {\bf 87 } 17007\\
8. Levy P M, Guo W and Cox D L, 1988 J. Appl. Phys. {\bf 63 }
3896\\
9. Pan S-H, O'Neal J P, Badzey R L, Chamon C, Ding H, Engeibrecht,
Wang Z, Eisaki H, Uchida S, Gupta A K, Ng K-W, Hudson
E W, Lang K M and Davis J C, 2001 Nature {\bf 413 } 282\\
10. Kaminski A, Rosenkranz S, Fretwell H M, Campuzano J C, Li Z,
Raffy H, Cullen W G, You H, Olson C G, Varma C M and Hochst H,
2002 Nature {\bf 416} 610\\
11. Guo W and Han R-S, 2001 Physica C {\bf 364-365} 79\\
12. Guo W and Han R-S, 2002 Chin. Phys. Lett. {\bf 19 } 1687\\
13. Tallon J L, Cooper J R, I.P.N. de-Silva P S, Williams G V M
and Loram J W, 1995 Phys. Rev. Lett. {\bf 75} 4114; Tallon J L,
Bernhard C, Shabed H, Hatterman R L and Jorgenson J D, 1995 Phys.
Rev. B {\bf 51} 12911
\end{document}